\documentclass[showpacs,twocolumn]{revtex4}

\usepackage{amsmath,amsthm,amscd,amssymb}
\usepackage{latexsym}

\usepackage{amsfonts}

\begin{document}

\title{\Large Constraining  Extended Theories of Gravity using Solar System Tests}
\date{\today}

\author{Gianluca ALLEMANDI}
\email{allemandi@dm.unito.it}
\affiliation{\footnotesize Dipartimento di Matematica, \\ Universit\`a di Torino, Via C. Alberto 10, 10123
Torino \\ and INFN, Sezione di Torino}
\author{Matteo Luca RUGGIERO}
\email{matteo.ruggiero@polito.it}
\affiliation{\footnotesize  Dipartimento di Fisica, \\ Politecnico di Torino,  Corso Duca degli Abruzzi 24, 10129 Torino \\ and INFN, Sezione di Torino}

\pacs{98.80.Jk, 04.50.+h, 04.20.-q}

\begin{abstract}
Solar System tests give nowadays constraints on the estimated value of the cosmological constant, which can be
accurately derived from different experiments regarding gravitational redshift, light deflection, gravitational
time-delay and geodesic precession. Assuming that each reasonable theory of gravitation should satisfy Solar
System tests, we use this limits on the estimated value of the cosmological constant to constrain  extended
theories of Gravity, which are nowadays studied as possible theories for  cosmological models and provide viable
solutions to the cosmological constant problem and the explanation of the present acceleration of the Universe.
We obtain that the estimated values, from Solar System tests, for the parameters appearing in the  extended
theories of Gravity  are orders of magnitude bigger than the values obtained in the framework of cosmologically
relevant theories.
\end{abstract}

\date{\today}

\maketitle

\section{Introduction}\label{sec:intro}

The fact that the Universe is now undergoing a phase of accelerated expansion is evidenced by many recent
observational tests, such as the light curves of the type Ia supernovae  and the cosmic microwave background
(CMB) experiments \cite{c3, c1, c11, d1, d2, d3}. The present acceleration of the Universe can be explained,
within General Relativity (GR), by claiming the existence of the \textit{dark energy}, a cosmic fluid having
exotic properties, since it is required to have equation of state such that $p<-\rho$, i.e. it has negative
pressure smaller than
its energy density. \\
Alternatively, it has been suggested that cosmic speed-up can be  explained by generalizing GR: introducing in
the gravitational action terms non linear in the scalar curvature $R$ (see \cite{c6,odintsov06} and references
therein). These theories are the so called $f(R)$ theories of gravity (or ``extended theories of gravity''), where the gravitational Lagrangian
depends on an arbitrary analytic function $f$ of the scalar curvature $R$. It was also shown recently that $f(R)$ gravity is a cosmologically viable theory because it may contain matter dominated and radiation phase before acceleration \cite{matty}.\\
In spite of the apparent difficulties
in agreeing with the recent cosmological observations, GR is in excellent agreement with the Solar System
experiments: hence, every theory that aims at explaining the accelerated expansion of the Universe, should
reproduce GR at the Solar System scale. Then, there is the need of checking $f(R)$ predictions with Solar System
tests or imposing constraints that derive from the experiments at Solar System scale. Actually, the debate is
still open and, since $f(R)$ theories can be studied both in the metric formalism
\cite{metfr1,metfr2,metfr3,metfr4} and in the  Palatini formalism
\cite{palfR1,palfR2,palfR3,palfR4,palfR5,palfR6,palfR7,palfR9}, the problem can be faced within both approaches
(see \cite{faraoni06} and references therein).

Actually, we must say that  there is not a common agreement on  the dynamics of $f(R)$ theories  in  presence of
matter, i.e. inside the sources of the gravitational field, both in the metric and in the Palatini formalisms
\cite{chiba03,flanagan03}.

Some recent papers state that $f(R)$ theories do not match Solar System Tests \cite{pitr,jin06,chiba06}. Indeed,
these analyses are carried out in the metric formalism and lead to the same results obtained some years ago by
Chiba \cite{chiba03}. For a different viewpoint, always in the metric formalism, which leads to an agreement
with Solar System Tests, see \cite{capozziello07,zhang07}. The difficulties arising in the metric formalism with the Solar
System experiments, which ultimately depend on the matching between the solutions inside and outside the matter
distribution, are not present in the Palatini formalism, as it has been recently showed \cite{kainulainen06}: in
fact, in this case the space-time inside a star does not affect the space-time outside it, (i.e.  the vacuum
solution), contrary to what happens in the metric formalism. Also, it is useful to remember that the solutions
of the Palatini field equations are a subset on solutions of the metric field equations: thus Palatini $f(R)$
theories differ strongly from metric $f(R)$ theories  theory \cite{magnano}.

In a previous work, we found an exact solution of the field equations in vacuum, in the Palatini approach  and
examined the significance of Post-Newtonian parameters thus arising \cite{allemandi05}. This solution
corresponds to the Schwarzschild-de Sitter metric, already studied in detail in \cite{zerbo}, and the
modifications to GR are related to the solutions of a scalar valued equation, that we called the
\textit{structural equation}. Owing to this fact,  we showed that the GR limit (i.e. the Schwarzschild solution)
and the Newtonian limit are recovered in the Solar System scale, thus suggesting the reliability of $f(R)$ at
this scale. On the other hand, the question arises of evaluating the constraints, coming to Solar System
experiments, on the $f(R)$ theories and, in particular, on the allowed analytical functions $f$ of the curvature
scalar. Thanks to the exact Schwarzschild-de Sitter solution,  we suggest here that this task can be, in
principle, accomplished, according to a procedure that has been introduced in \cite{ruggiero06}. In fact, recent
papers by \cite{kagramanova06,sereno06a,iorio06a} regarding the constraints on the Schwarzschild-de Sitter
metric coming for the Solar System experiments, provide a method to give the best estimate for the parameter $k$
which, in our approach, is a measure of the non-linearity of the theory. What we ultimately aim at is obtaining
constraints on the parameters appearing in the $f(R)$ functions, that is to say on the allowed forms of $f(R)$.
As a by product, since our approach lends itself to set limits on the allowable values of the scalar curvature
$R$, we may infer corresponding bounds on the cosmological density of (visible) matter.

\section{Constraints on $f(R)$ from Solar System experiments}\label{sec:cons}

The field equations of $f(R)$ theories, in the Palatini formalism, which have the explicit form:
\begin{eqnarray}
f^{\prime }(R) R_{(\mu\nu)}(\Gamma)-\frac{1}{2} f(R)  g_{\mu \nu
}&=&\kappa T_{\mu \nu }^{mat}  \label{ffv1}\\
\nabla _{\alpha }^{\Gamma }[ \sqrt{g} f^\prime (R) g^{\mu \nu })&=&0
\label{ffv2}
\end{eqnarray}
admit the spherically symmetric vacuum exact solution:
\begin{align}
ds^2=&-\left(1-\frac{2M}{r}+\frac{k r^2}{3}\right)dt^2+\frac{dr^2}{\left(1-\frac{2M}{r}+\frac{k r^2}{3}\right)}
\notag \\ & +r^2d\vartheta^2+r^2\sin^2 \vartheta d\varphi^2, \label{eq:metrica1}
\end{align}
which is commonly referred to as Schwarzschild-de Sitter metric \footnote{Our  results hold if the Lagrangian is
not in the form $f(R)=R^n$, with $n \geq 2, n \in \mathbb{N}$.}: we refer to \cite{allemandi05}  for  details on
the derivation of this solution. The parameters appearing in (\ref{eq:metrica1}) are the mass $M$ of the
spherically symmetric source of the gravitational field and $k$, which is related to the solutions $R=c_i$ of
the structural equation
\begin{equation}
f^{\prime }(R) R-2f(R)= 0. \label{eq:struct1}
\end{equation}
controlling the solutions of equation (\ref{ffv1}).
Namely, it is $k=c_i/4=R/4$. We point out that (\ref{eq:metrica1}) is a constant curvature solution. Indeed, we
may say that $k$ is a measure of the non-linearity of the theory (if $f(R)=R$, eq. (\ref{eq:struct1}) has the
solution $R=0 \rightarrow k=0$). It is evident, at least from a theoretical viewpoint, that the $f(R)$ contribution to the gravitational potential should be small enough not to contradict the known tests of gravity. In the cases of small values of $R$ (which
surely occur at Solar System scale) the Einsteinian limit (i.e. the Schwarzschild solution) and the Newtonian
limit are recovered, as it is manifest from (\ref{eq:metrica1}).\\
On the other hand, inspection of the Schwarzschild-de Sitter metric, suggests that, when $M=0$ the (constant
curvature) solutions are not, in general, Minkowski space-time, but they  should be de Sitter or anti de Sitter space-time:
as a matter of facts, Minkowski space-time is solution of the field equations only if $R=0$ (see \cite{allemandi05}). \\

In GR, the Schwarzschild-de Sitter solution corresponds to a spherically symmetric solution of Einstein
equations, with a "cosmological" term $\Lambda g_{\mu\nu}$, $\Lambda$ being the cosmological constant, which
might be, in turn, one of the candidates for explaining the accelerated expansion of the Universe, even though
the nature of the cosmological constant is still under debate \cite{peebles03}. In practice, it is $\Lambda=-k$
in our notation. A recent paper by Kagramanova \textit{et al.} \cite{kagramanova06} calculates the classical
gravitational effects in the Solar System on the basis of the Schwarzschild-de Sitter metric. On studying
the gravitational redshift, the deflection of light, the gravitational time-delay and the geodetic precession, constraints are
deduced on the magnitude of the cosmological constant. Besides the deflection of light, which is not influenced by
$\Lambda$, the other tests give estimates on $\Lambda$ that are much greater than those estimated from
observational cosmology. Indeed, these estimates range from $|\Lambda| \leq 10^{-24} \ m^{-2}$ to $|\Lambda|
\leq 10^{-27} \ m^{-2}$ while, in cosmology, the current value of the cosmological constant is estimated to be
$\Lambda_0 \simeq 10 ^{-52} m^{-2}$ \cite{peebles03}.

Using a different approach, Jetzer and Sereno \cite{sereno06a}  suggest that the best constraint $\Lambda_0
\simeq 10 ^{-42} m^{-2}$ comes from the extra precession on the basis of Earth and Mars observations. A similar
result was also obtained by Iorio \cite{iorio06a}.

Armed with this results, we may reverse the argument, and obtain, from these constraints on $|\Lambda|=|k|$,
limits on the $f(R)$ functions used in cosmology to explain cosmic speed-up. This is a further test to obtain constraints on $f(R)$ gravitational theories and to test their viability with well accepted experimental tests for Gravity.

As we have seen, the $k$ parameter is simply related to the solutions of the
structural equation
\begin{equation}
f^{\prime }(R) R-2f(R)= 0, \label{eq:struct12}
\end{equation}
namely it is $k=c_i/4$, where $R=c_i$ are the solutions of (\ref{eq:struct12}). As a consequence, the bounds on
the $k$ parameter, in principle,  enable us to constrain the functions $f(R)$. We may proceed as follows.
In general, the functions $f$, beyond the scalar curvature $R$, depend on a set of $N$ constant parameters
$\alpha_j$, $j=1..N$, so that we may write $f=f(R,\alpha_1,..,\alpha_N)$, and, on solving (\ref{eq:struct12}), we
obtain
\begin{equation}
R=\mathcal{F}(\alpha_1,..,\alpha_N). \label{eq:falpha1}
\end{equation}

Consequently, what we ultimately obtain is a limit on the allowed values of the combination $\mathcal{F}$ of
these parameters.

For instance, let us considered the Lagrangian introduced in \cite{c6}, which mimics cosmic acceleration without
need for dark energy:
\begin{equation}
f(R)=R-\frac{\mu^4}{R}. \label{eq:frcarrol1}
\end{equation}
We point out that the Lagrangian (\ref{eq:frcarrol1}) was found to be instable \cite{metfr1,metfr2,dolgov03}. For
discussion on this issue, we refer to the recent paper by Faraoni \cite{faraoni06b}. Nonetheless, we consider
the Lagrangian (\ref{eq:frcarrol1}) as a first  example in order to illustrate our approach.

The Lagrangian  (\ref{eq:frcarrol1}) depends on the parameter $\mu$ only, so that  eq. (\ref{eq:falpha1})
becomes
\begin{equation}
|R|=\sqrt{3}\mu^2,  \label{eq:frcarrol2}
\end{equation}
and, consequently, we may set a limit on the parameter $\mu$. If we take the value $|\Lambda| \leq \times
10^{-42} m^{-2}$ from \cite{sereno06a}, we get
\begin{equation}
\mu \leq   10 ^ {-21} m^{-1}, \label{eq:carrol3}
\end{equation}
or, since $\mu$ is a parameter whose physical units are those of a mass,
\begin{equation}
\mu \leq 10 ^ {-28} eV. \label{eq:carrol4}
\end{equation}
We can alternatively use the best estimate $|\Lambda| \leq 10^{-27} \ m^{-2}$ given in \cite{kagramanova06} and we correspondingly
get
\begin{equation}
\mu \leq   10 ^ {-20} eV. \label{eq:carrol5}
\end{equation}
We notice that both values (\ref{eq:carrol4},\ref{eq:carrol5}) are remarkably greater than estimate $\mu \simeq
10 ^ {-33} \ eV$ \cite{c6}, needed for $f(R)$ gravity to explain the acceleration of the Universe without
requiring dark matter.

A most significant example is given  by polynomial-like functions $f(R)$; it can be showed that the structural
equation can be solved analytically. This is the case already examined in \cite{palfR2}:
\begin{equation}
f(R)=\alpha R + \frac{\beta}{m-2}R^m (m \neq 1,2), \label{ea:alle1}
\end{equation}
from which we obtain, on solving the structural equation:
\begin{equation}
|\mathcal{F} \left(\alpha,\beta,m \right)|= \left|\frac{\alpha}{\beta}^{\frac{1}{1-m}}\right| ,\label{eq:alle2}
\end{equation}
This implies that the same bounds can be derived on the function $\mathcal{F} \left(\alpha,\beta,m \right)$:
\begin{equation}
|\mathcal{F} \left(\alpha,\beta,m \right)|  \leq   10 ^ {-42} m^{-2},  \label{eq:al1}
\end{equation}
or
\begin{equation}
|\mathcal{F} \left(\alpha,\beta,m \right)|  \leq   10 ^ {-27} m^{-2},  \label{eq:al2}
\end{equation}
according to  \cite{sereno06a} or \cite{kagramanova06} respectively. This example is relevant from a
cosmological viewpoint: the present form of $f(R)$ reproduces an inflationary behavior of the model, in the case
$m>0$ and, on the contrary, the case $m<0$ is important to construct and study accelerated models of the present
Universe as previously remarked.

We can also consider the Lagrangian \cite{palfR1,palfR2}
\begin{equation}
f(R)=R-\frac{6\alpha}{\sinh R/\alpha}, \label{eq:alle3}
\end{equation}
in the limit of small $R$, such that the structural equation gives
\begin{equation}
|R|=3\alpha \label{eq:alle4}
\end{equation}
Consequently, we may set limits on the parameter $\alpha$, in particular
\begin{equation}
\alpha \leq 10^{-42} m^{-2}, \label{eq:alle5}
\end{equation}
and
\begin{equation}
\alpha \leq 10^{-27} m^{-2}, \label{eq:alle6}
\end{equation}
following to \cite{sereno06a}, or \cite{kagramanova06}, respectively. According to \cite{palfR2}, we must take
$\alpha \simeq 10^{-52} m^{-2}$ to match observations of the cosmological acceleration, and, once again, the
cosmologically relevant values are much smaller than those obtained from Solar System experiments.

What we have seen suggests that the constraints deriving from Solar System tests are surely coherent with
cosmological theories in the framework of  extended theories of Gravity, which are able to explain the present
acceleration of the Universe. The feeling is that data coming from one context cannot be used to infer anything
on the other context, since  they have very different orders of magnitude.

The dichotomy between Solar System experiments and cosmological observations can be seen from a different
viewpoint. As we have said, $f(R)$ theories of gravity are expedient to explain cosmic speed-up, thanks to a
geometrical approach, without requiring the existence of dark energy or other exotic fluids. This can be
accomplished by studying the Friedman-Robertson-Walker (FRW) cosmology in $f(R)$ gravity \cite{palfR2}. In this case, the structural
equation (\ref{eq:struct1}) reads
\begin{equation}
f^{\prime }(R) R-2f(R)= \kappa \tau, \label{eq:struct11}
\end{equation}
where $\tau=\mathrm{tr} T=g^{\mu\nu}T_{\mu\nu}^{mat}$, $T_{\mu\nu}^{mat}$ being the stress-energy tensor of the
cosmological fluid. Re-phrasing the approach outlined above, the solution of eq.
(\ref{eq:struct11}) can be written in the form
\begin{equation}
R=\mathcal{G}(\alpha_1,..,\alpha_N,\tau). \label{eq:galphatau1}
\end{equation}
In other words, what we can get is a limit on the allowed values of the combination $\mathcal{G}$ of the
$\alpha_j$ parameters and the trace $\tau$ of the stress-energy tensor of the cosmological fluid.

For instance, from the Lagrangian (\ref{eq:frcarrol1}), we obtain
\begin{equation}
R=\mathcal{G}(\mu,\tau)= \frac{-\tau\pm \sqrt{\tau^2+12\mu^4}}{2}. \label{eq:solrmutau1}
\end{equation}
If we consider the case of dust-dominated Universe, $\tau=-\rho$: as a consequence, (\ref{eq:solrmutau1}) allows
to get information on the allowed values of the combination of the matter density $\rho$ and the parameter
$\mu$. Actually, since the parameter $\mu$ is required to be very small in order to match the cosmic speed-up
\cite{palfR1}, the relation (\ref{eq:solrmutau1}) sets limits on the values of the (visible) matter density
$\rho$, which, thanks to the data reported by \cite{sereno06a} or \cite{kagramanova06}, are easily seen to be
much different from those deriving from cosmological observations.

\section{Discussions and Conclusions}\label{sec:disconc}

The recent estimates of $k$ (or $\Lambda$) from Solar System experiments suggest some comments on the
reliability of  extended theories of gravity.

First of all, up today,   no observation within the Solar System is able to reveal effects due to the non linearity of
$f(R)$ (or, rephrased in GR language, due to the cosmological constant). Strictly speaking,  this statement is the outcome of our analysis that has been carried in the Palatini formalism of $f(R)$ gravity; however similar conclusions seem to arise also in the metric  approach, as recent papers
suggest (see \cite{capozziello07,zhang07} and references therein).

Second, conversely, present
observations at the Solar System scale are hardly fit to constrain the forms of the $f(R)$ functions used in
cosmology, setting bounds on the parameters appearing in them, since the Solar System estimates of these
parameters are much greater than the cosmologically required values. This is confirmed also by estimating the
density of visible matter, which turns out to be several orders of magnitude different from the one derived
thanks to cosmological observations.

Of course, these tests are done on a particular solution of the $f(R)$ equations, namely a spherically symmetric
vacuum solution, corresponding to constant scalar curvature. Even if relaxing the spherical symmetry does not
seem to produce relevant changes, different conclusions might arise considering the solutions inside a matter
distribution, such as a galaxy, thus enlarging the scale under consideration.

The underlying feeling is that non-linearities appearing in $f(R)$ become important on scales much larger than
the Solar System, for sure on the cosmological scale and, hence, data coming from one context cannot be used to
infer anything on the other context: in fact, they belong to a different realm of application of the same
theory, if one trusts in $f(R)$, or to different approximations of a more general theory, if one believes that
GR at the Solar System scale and $f(R)$ in cosmology are just good approximations of something else, still
undefined.

\section*{ACKNOWLEDGMENTS}
M.L.R  acknowledges financial support from the Italian Ministry of University and Research (MIUR) under the
national program ``Cofin 2005'' - \textit{La pulsar doppia e oltre: verso una nuova era della ricerca sulle
pulsar}.\\


\end{document}